\documentclass[12pt, a4paper, twoside]{fec}
\usepackage{graphicx}
\usepackage{bm}
\usepackage{amsmath}             

\newcommand{\ee}{&\hspace{-.5em}}
\newcommand{\sub}[1]{\hspace{-.07em}\bm{#1}}

\newcommand{\mtr}[1]{\mathrm{#1} }
\newcommand{\para}{\parallel }

\newcommand{\pdif}[2]{\ensuremath{\frac{\partial #1}{\partial #2}}}

\newcommand{\D}{\cdot}

\newcommand{\lbr}{\left ( }
\newcommand{\rbr}{\right ) }

\newcommand{\vpr}{v_{\parallel} }
\newcommand{\vpp}{v_{\perp} }
\newcommand{\vv}{v}
\newcommand{\bb }{\bm{b} }

\newcommand{\dfg }{\delta f^{\mathrm{(g)}} }

\newcommand{\dphik }{\delta \phi_{\sub{k}_{\perp}} }
\newcommand{\dalk }{\delta A_{\parallel \sub{k}_{\perp}} }
\newcommand{\dpsi }{\delta \psi }
\newcommand{\dpsik }{\delta \psi_{\mtr{s}\sub{k}_{\perp} }}

\newcommand{\mr}[1]{\text{#1}}

\title{Gyrokinetic turbulent transport simulations on steady burning condition in D-T-He plasmas}

\author[1,2,3]{$^{\#}$Motoki Nakata}
\author[4]{Mitsuru Honda}
\affil[1]{National Institute for Fusion Science, National Institutes of Natural Sciences, Toki 509-5292, Japan}
\affil[2]{The Graduate University for Advanced Studies, Toki 509-5292, Japan}
\affil[3]{PRESTO, Japan Science and Technology Agency, 418, Honcho, Kawaguchishi, Saitama 332-0012, Japan}
\affil[4]{Graduate School of Engineering, Kyoto University, Nishikyo, Kyoto 615-8530, Japan}

\doi

\begin{document}

\maketitle

\setstretch{1.2}

\begin{abstract}
{
Ion temperature gradient(ITG) and trapped electron modes(TEM) driven turbulent transport in an ITER-like plasma 
is investigated by means of multi-species gyrokinetic Vlasov simulations with D, T, He, and real-mass kinetic 
electrons including their inter-species collisions. 
Beyond the conventional zero-dimensional power balance analysis presuming the global energy and particle confinement times, 
gyrokinetic-simulation-based evaluation of a steady burning condition with He-ash exhaust 
and D-T fuel inward pinch is demonstrated. 
It is clarified that a significant imbalance appears in the turbulent particle flux for the fuel ions of D and T, 
depending on the D-T density ratio and the He-ash accumulation. 
Then several profile regimes to satisfy Reiter's steady burning condition are, for the first time, identified by the gyrokinetic simulation.  
Also, the impacts of zonal flows and nonthermal He-ash on the optimal profile regimes are examined. 
\ \\

\small{\textbf{This is the author accepted manuscript (AAM/AM), published in Plasma Fusion Research 17, 1403083 (2022)\ \  doi: 10.1585/pfr.17.1403083 \\
\ (with permission from JSPF)}
}
\ \\ \ \\ \ \\
\footnotesize{$\#\,$Present affiliation: Faculty of Arts and Sciences, Komazawa University, Tokyo 154-8525, Japan}
}
\end{abstract}

\newpage 

\section{Introduction}
\ \ \ \ \ Burning plasmas are intrinsically composed of multi-ion species such as fuel hydrogen isotopes(Deuterium and Tritium), 
the high-energy $\alpha$ particles produced by the fusion reaction, the thermalized He-ash, 
and high-Z impurities resulting from the plasma-facing materials. 
More complex turbulent transport processes are, thus, expected in the burning plasmas, 
compared to those in the single or very few ion-species plasmas, which are usually addressed in experiments. 
Since simultaneous measurements of the kinetic profiles for all particle species including the inter-species energy exchange are difficult, 
the first-principle-based gyrokinetic turbulent transport simulations for the multi-ion mixture plasmas are indispensable 
to predict the confinement performance, and to optimize the D-T fueling and the impurity exhausts in the burning plasma. 
There are several earlier works regarding the ITG driven turbulent transport 
of fuel hydrogen-isotope ions\cite{estrada1,estrada2,nakata_prl,belli,Bourdelle} 
and the light and heavy impurity transport\cite{angioni1,angioni2}, 
but the fuel and impurity ion transport has often been examined separately. 
The impacts of multi-ion species on the ion- and electron-scale microinstabilities have also been studied\cite{nakata_pfr2022}, 
and the reduction of turbulent heat flux through multi-scale interactions in the multi-ion species plasma has been revealed\cite{maeyama}. 
In addition, several recent experiments in the mixture\cite{ida1,ida2,urano} and/or the D-T burning plasmas\cite{Mailloux} 
with dedicated controls of the ion density have revealed the significant isotope- and impurity-effects 
on turbulent particle and energy transport, which are important for optimizing the fusion power output. 

Multi-ion species not only complicates the transport processes in burning plasmas, 
but also strongly affects the ignition and steady burning conditions. 
Indeed the zero-dimensional power balance analysis shows that 
the accumulation of the thermalized He-ash in the core region leads to a more severe self-ignition criteria 
on the Lawson diagram, as shown in Fig. 1.  
Here the ignition curves on a plane of the fusion triple product $n_{0}\hat{\tau}_{\mr{E}}T_{0}$ and the on-axis temperature $T_{0}$ 
are calculated according to Ref. 14 and 15, 
where $n_{0}$ and $\hat{\tau}_{\mr{E}} = (1+P_{\mr{rad}}/P_{\mr{abs}})^{-1}\tau_{\mr{E}}$ 
denote the on-axis density and the effective global energy confinement time including the radiation-loss power $P_{\mr{rad}}$. 
%
%
In addition to the ignition curves, the steady burning condition is described by the closed curves, 
where the ratio of the particle(He-ash) to energy confinement time, 
i.e., $\rho_{\tau} \! := \! \tau_{\mr{p}}/\tau_{\mr{E}}$, where $\tau_{p} \! \simeq \! \tau_{\mr{He}}$, is the critical parameter, 
as will be discussed later.  
It should be noted, however, that any transport processes of the fuel ions and He-ash are not incorporated 
in the above argument. 
The multi-species turbulent transport studies are, thus, required to examine the steady burning condition, 
beyond the conventional zero-dimensional power balance analysis. 

In this study, the multi-species gyrokinetic Vlasov simulations have been carried out 
to clarify the profile conditions with optimal balance of turbulent energy and particle transport in the steady burning state. 
Here, Reiter's steady burning condition $\tau_{\mr{He}} < \alpha^{\ast} \tau_{\mr{E}}$\cite{reiter}, 
which describes a permissible ratio of the He-ash confinement to the energy confinement, is investigated, 
where $\alpha^{\ast} \! \sim \!$ 7--15 depending on the wall and divertor conditions. 
We consider the ion temperature gradient(ITG) and trapped electron mode(TEM) driven turbulent transport 
in an ITER-like plasma\cite{nakata_pfr2014} with D, T, the thermalized He-ash, and real-mass kinetic electrons 
including their inter-species collisions\cite{nakata_cpc2015}. 
A good prediction capability of the gyrokinetic simulation has been confirmed for the JT-60U tokamak L-mode experiment\cite{nakata_nf2016}. 
The rest of the paper is organized as follows. 
In Sec. 2, the gyrokinetic simulation model and the numerical setup are described. 
Then the linear and nonlinear ITG-TEM turbulence simulation results are presented. 
In Sec. 3, we discuss the impact of the D-T ratio and the He-ash accumulation on the energy and particle transport. 
The gyrokinetic-simulation-based evaluation of the profile regimes to satisfy the Reiter's steady burning condition are presented in Sec. 4.    
Finally, the summary is given in Sec. 5. 
%
%
\section{Multi-species gyrokinetic turbulence simulations}
\subsection{Gyrokinetic model}
\ \ \ \ In this study, a multi-species electromagnetic gyrokinetic Vlasov solver GKV\cite{gkv} is used. 
The simulation model is briefly summarized in this section. 
The governing equation is an electromagnetic gyrokinetic equation for the particle species ``s'', 
which describes the time evolution of the non-adiabatic part of the perturbed gyrocenter distribution function 
$\delta g_{\mr{s}}$ in the five-dimensional phase-space ($\bm{x}_{\mr{g}},\vpr,\mu$). 
The Fourier representation with respect to the perpendicular wavenumber $\bm{k}_{\perp}$ is given by
%
\begin{eqnarray} 
\left ( \pdif{}{t} + \vpr \bb \! \D \! \nabla + i\omega_{\mr{Ds}} - \frac{\mu\bb \! \D \! \nabla B}{m_{\mr{s}}}\pdif{}{\vpr}
\right ) \delta g_{\mr{s}\sub{k}_{\perp}}  
- \frac{c}{B} \! \sum_{\Delta} \bb \! \D \! \lbr \bm{k}_{\perp}^{\prime} \! \times \! \bm{k}_{\perp}^{\prime \prime} \rbr 
\dpsi_{\mr{s}\sub{k}_{\perp}^{\prime}} \delta g_{\mr{s} \sub{k}_{\perp}^{\prime \prime}} \nonumber \\
 =  \frac{e_{\mr{s}}F_{\mr{Ms}}}{T_{\mr{s}}}\left (
\pdif{}{t} 
+ i\omega_{\ast T\mr{s}} 
+ \vpr \frac{\mu\bb \! \cdot \! \nabla B}{T_{\mr{s}}}
  \right ) \dpsik 
+ \mathcal{C}_{\mr{s}} \! \left ( \delta g_{\mr{s}\sub{k}_{\perp}} \right )\ ,
\end{eqnarray}
%
where $\delta g_{\mr{s}\sub{k}_{\perp}} \! = \! \dfg_{\mr{s}\sub{k}_{\perp}} + F_{\mr{Ms}}J_{0\mr{s}}e_{\mr{s}}\dphik/T_{\mr{s}}$, 
and $J_{0\mr{s}} \! = \! J_0(k_{\perp}\vpp/\Omega_{\mr{s}})$ is the zeroth-order Bessel function. 
The gyrophase-averaged electrostatic and electromagnetic potential fluctuations are combined into 
$\dpsik \! = \! J_{0\mr{s}}\dphik \! - (\vpr/c)J_{0\mr{s}}\dalk$. 
Since we focus on the finite but low-$\beta$ plasmas here, the parallel magnetic field fluctuation
$\delta B_{\para \sub{k}_{\perp}}$ is ignored.  
The equilibrium part of the distribution function is given by the local Maxwellian distribution, i.e., 
$F_{\mr{Ms}} \! = \! n_{\mr{s}}(m_{\mr{s}}/2\pi T_{\mr{s}})^{3/2}\exp[-(m_{\mr{s}}\vpr^{2}+2\mu B)/2T_{\mr{s}}]$.

The gyrokinetic equation shown in Eq. (1) is solved in the local fluxtube coordinates $(x,y,z)$ 
defined as $x=a(\rho-\rho_0),\ \, y=a\rho_{0}q(\rho_{0})^{-1}\left [ q(\rho)\theta - \zeta \right ],\ \, z=\theta$ 
with the straight-field-line flux coordinates $(\rho,\theta,\zeta)$, where $a$ and $q(\rho_{0})$ 
denote the plasma minor radius and the safety factor on the flux surface of interest $\rho_{0}$, respectively. 
In these coordinates, the magnetic and diamagnetic drift frequencies for $\bm{k}_{\perp} \! = \! k_x \nabla x + k_y \nabla y$ are given by 
%
%
\begin{eqnarray} 
\omega_{\mtr{Ds}} \ee = \ee  \frac{cT_{\mtr{s}}}{e_{\mtr{s}}B}\bm{k}_{\perp} \! \D \bb \times 
\lbr \mu \nabla B + m_{\mtr{s}}\vpr^{2}\bb \! \D \! \nabla \bb \rbr \nonumber \\ 
\ee = \ee \frac{c ( m_{\mtr{s}}\vpr^2 \! +\! \mu B )}{e_{\mtr{s}}B_{\mtr{ax}}}\lbr 
\mathcal{K}_{\mr{g}}k_{x}+\mathcal{K}_{\mr{n}}k_{y} \rbr \ , \\
\omega_{\ast T\mtr{s}} \ee = \ee  \frac{cT_{\mtr{s}}}{e_{\mtr{s}}B}\left [ 
1  + \eta_{\mtr{s}}\lbr \frac{m_{\mtr{s}}\vpr^{2} \! + \! 2\mu B}{2T_{\mtr{s}}}-\frac{3}{2} \rbr \right ] 
\bm{k}_{\perp} \! \D \bb \times \nabla \ln n_{\mtr{s}}\ ,
\end{eqnarray}
%
where $\eta_{\mtr{s}}=L_{n_{\mtr{s}}}/L_{T_{\mtr{s}}}$ with $L_{n_{\mtr{s}}}=-(d\ln n_{\mtr{s}}/dx)^{-1}$ and 
$L_{T_{\mtr{s}}}=-(d\ln T_{\mtr{s}}/dx)^{-1}$. 
The geometric coefficients representing the geodesic curvature component $\mathcal{K}_{\mr{g}}$ 
and the normal curvature component $\mathcal{K}_{\mr{n}}$ are defined as follows:
%
\begin{eqnarray} 
\mathcal{K}_{\mr{g}} = \frac{g^{xz}g^{xy}-g^{xx}g^{yz}}{B^2/B_{\mtr{ax}}^2}\pdif{\ln B}{z} -\pdif{\ln B}{y} \ , \\
\mathcal{K}_{\mr{n}} = \frac{g^{xz}g^{yy}-g^{xy}g^{yz}}{B^2/B_{\mtr{ax}}^2}\pdif{\ln B}{z} +\pdif{\ln B}{x} \ , 
\end{eqnarray}
%
where $g^{ij}, (i,j) \! =\! (x,y,z)$ denotes the contravariant components of the metric tensor. 
Note that the pressure gradient effect in the curvature drift is ignored by the low-$\beta$ approximation, 
and $\partial \ln B/\partial y = 0$ for the axisymmetric fields. 

Collisional effects are introduced in terms of a multi-species gyrokinetic collision operator $\mathcal{C}_{\mr{s}}$ expressed as 
%
\begin{equation}
\mathcal{C}_{\mr{s}}(\delta g_{\mr{s}\sub{k}_{\perp}}) = 
\sum_{\mr{s}^{\prime}}\oint \frac{d\varphi}{2\pi}e^{i\bm{k}_{\perp}\D\bm{\rho}_{\mr{s}}} \left \{ 
\mathcal{C}^{\mr{T}}_{\mr{ss}^{\prime}}[e^{-i\bm{k}_{\perp}\D\bm{\rho}_{\mr{s}}}\delta g_{\rm{s}\bm{k}_{\perp}}] 
+\mathcal{C}^{\mr{F}}_{\mr{ss}^{\prime}}[e^{-i\bm{k}_{\perp}\D\bm{\rho}_{\mr{s}^{\prime}}}\delta g_{\rm{s}^{\prime}\bm{k}_{\perp}}] \right \}, 
\end{equation}   
%
where the field-particle part $\mathcal{C}^{\mr{F}}_{\mr{ss}^{\prime}}$ holds the particle, momentum, and energy conservation, 
as well as Boltzmann's H-theorem\cite{nakata_cpc2015,sugama_pop2009}. 

The electromagnetic fluctuations are determined by the Poisson-Amp\`ere equations: 
%
\begin{eqnarray}
\lbr k_{\perp}^{2} \! +\lambda_{\mr{D}}^{-2} \rbr \dphik  = 4\pi \sum_{\mr{s}} e_{\mr{s}} \! \! \int \! \! 
d\bm{\vv} J_{0\mr{s}}\delta g_{\mr{s}\sub{k}_{\perp}} \ , \\
k_{\perp}^{2} \dalk  = \frac{4\pi}{c} \sum_{\mr{s}} e_{\mr{s}} \! \! \int \! \! 
d\bm{\vv}\, \vpr J_{0\mr{s}}\delta g_{\mr{s}\sub{k}_{\perp}} \ , 
\end{eqnarray}   
%
%
where $\lambda_{\mr{D}} = (\sum_{\mr{s}}4\pi n_{\mr{s}} e_{\mr{s}}^2/T_{\mr{s}})^{-1/2}$ is the Debye length. 
The charge neutrality in the background density $n_{\mtr{s}}$ is described as $\sum_{\mtr{s}\neq \mtr{e}}f_{C\mtr{s}}\! = \! 1$,
where $f_{C\mtr{s}} \! \equiv \! Z_{\mtr{s}}n_{\mtr{s}}/n_{\mtr{e}}$ means the charge-density fraction for ions
with the charge number $Z_{\mtr{s}}$.
The contribution of each species to the source terms in the right hand side of Eqs. (7) and (8) is proportional to $f_{C\mtr{s}}$.
In addition, the radial derivative of the above charge neutrality condition leads to another constraint for the background density 
gradient, i.e., $\sum_{\mtr{s} \neq \mtr{e} } f_{C\mtr{s}}L_{n_{\mtr{s}}}^{-1} \! = \! L_{n_{\mtr{e}}}^{-1}$.

The radial turbulent fluxes with respect to the energy and particle are, respectively, defined by 
%
\begin{eqnarray}
Q_{\mr{s}} \! :=\! \left < \bm{Q}_{\mr{s}} \! \cdot \! \nabla x \right > \! = \! 
\left < \frac{c}{B}\! \int \! d\bm{\vv}\,\frac{m_{\mr{s}}\vv^2}{2}  \lbr \nabla x \! \cdot \! \bb \! \times \! \nabla_{\perp} \delta \phi \rbr
\delta g_{\mr{s}} \right >, \\
\Gamma_{\mr{s}} \! :=\! \left < \bm{\Gamma}_{\mr{s}} \! \cdot \! \nabla x \right > \! = \!  
\left < \frac{c}{B}\! \int \! d\bm{\vv}\, \lbr \nabla x \! \cdot \! \bb \! \times \! \nabla_{\perp} \delta \phi \rbr \delta g_{\mr{s}} \right >, 
\end{eqnarray}   
%
where the flux-surface average is denoted by $\langle \cdots \rangle$. 
By using the above gyrokinetic Vlasov and Poisson-Amp\`ere equations, 
one can derive the entropy balance/transfer equation 
which describes the relation among the fluctuations of the phase-space distribution function, the transport fluxes, 
the entropy transfer via nonlinear interactions, and the collisional dissipation\cite{sugama_pop2009, nakata_pop2012}.   
It also provides us with a good measure for the accuracy of the kinetic turbulence simulation. 
%
%
\subsection{ITG-TEM driven turbulence simulations in ITER-like D-T-He plasma}
\ \ \ \ In this section, we present multi-species gyrokinetic simulations 
for the ITG-TEM instability and the associated turbulent transport in an ITER-like shaped plasma\cite{nakata_pfr2014}, 
where 4-species of D($A_{\mr{D}}\! =\! 2,Z_{\mr{D}}\! = \!1$), T($A_{\mr{T}}\! =\! 3,Z_{\mr{T}}\! = \!1$), 
He($A_{\mr{He}}\! =\! 4,Z_{\mr{He}}\! = \!2$), and real-mass kinetic electrons are incorporated. 
%
As a likely profile condition, we assume the normalized temperature and density gradient lengths at $\rho\! = \! 0.5$ 
to be $R_{\rm{ax}}/L_{T_{\rm{i}}}\! =\! 6$ for all the ion-species, $R_{\rm{ax}}/L_{T_{\rm{e}}}\! =\! 8$, 
and $R_{\rm{ax}}/L_{n} \! =\! 2$ for all the particle species, respectively. 
Also, the equal temperature of $T_{\mr{D}} \! = \! T_{\mr{T}} \! = \! T_{\mr{He}} \! = \! T_{\mr{e}} \! = \! 5\mr{keV}$, 
and the electron density of $n_{\mr{e}}=1.4 \! \times \! 10^{20}\mr{m}^{-3}$ 
are considered, where the normalized electron-electron collisionality is $\nu_{\mr{ee}}^{\ast} \sim 0.02$. 
As for the numerical resolution, (169, 49)-mode numbers in $(k_x,k_y)$ and (64, 64, 24)-grid numbers in $(z, \vpr, \vpp)$ 
are used, where $(k_{x}^{\mr{min}}\rho_{\mr{tH}},\ k_{y}^{\mr{min}}\rho_{\mr{tH}}) \! = \! (0.049,\ 0.05)$ 
and the velocity-space domain is taken with 
$(|\vpr^{\mr{max}}|,\, \vpp^{\mr{max}}) = (4.5\vv_{\mr{ts}},\, 4.5\vv_{\mr{ts}})$ at $z \! =\! 0$. 
A small but finite $\beta$ value of $\beta_{0} \! := \! n_{\mr{e}}T_{\mr{i}}/B_{\mr{ax}}^2 \! = \! 1 \! \times \! 10^{-3}$ 
is imposed to suppress the fast dispersive Alf\'{v}en waves so-called the $\omega_{\mr{H}}$ mode. 

The linear growth rate and mode frequency in the D-T-He plasma with $10\%$ He-ash, 
i.e., $n_{\mr{D}}/n_{\mr{e}} \! =\! n_{\mr{T}}/n_{\mr{e}} \! =\! 0.4$ and $n_{\mr{He}}/n_{\mr{e}} \! =\! 0.1$, 
are shown in Fig. 2, 
where the cases with pure-D, pure-T, and D-T are also plotted for comparisons. 
Note that the effective thermal speed $v_{\mr{t(eff)}}$ and ion gyroradius $\rho_{\mr{t(eff)}}$ 
are defined by using the effective ion mass and charge numbers, 
i.e., $A_{\mr{eff}} \! \equiv \! \sum_{\mr{s} \ne \mr{e}}f_{C\mr{s}}A_{\mr{s}}$ 
and $Z_{\mr{eff}} \! \equiv \! \sum_{\mr{s}\ne \mr{e}}f_{C\mr{s}}Z_{\mr{s}}$. 
One can see a only slight change of the normalized growth rate and the mode frequency even in the case with He-ash accumulation, 
where the dilution effect is absorbed in the normalization with the charge-density fraction $f_{\mr{Cs}}$. 

The time evolution of turbulent energy and particle fluxes, i.e., $Q_{\mr{s}}$ and $Z_{\mr{s}}\Gamma_{\mr{s}}$, 
in the D-T-He plasma with $n_{\mr{D}}\! =\! n_{\mr{T}} \! =\! 0.4n_{\mr{e}}$ and $n_{\mr{He}}\! =\! 0.1n_{\mr{e}}$ is shown in Figs. 3(a) and 3(b), 
where the gyro-Bohm unit for hydrogen is used here.  
Depending on the particle species, different saturation levels, e.g., $Q_{\mr{He}} < Q_{\mr{T}} \sim  Q_{\mr{D}} \sim 0.5Q_{\mr{e}}$ 
appear in the statistically steady state, 
where a slight deviation from the gyro-Bohm scaling is found for the hydrogen isotopes of D and T. 
It has been confirmed that the ambipolar condition of $\Gamma_{\mr{D}}  + \Gamma_{\mr{T}}  + 2\Gamma_{\mr{He}}  =  \Gamma_{\mr{e}}$
is accurately satisfied. 
%
Note also that, in contrast to the quasilinear flux,  
the direction of the T particle transport flips from inward to outward after the nonlinear saturation 
of the linear growth($t \! \sim \! 45 R_{\mr{ax}}/\vv_{\mr{tH}}$). 
The snapshot of the temperature fluctuations in the real space is displayed in Fig. 3(c), 
where the fluctuations of D, T, and He are simultaneously visualized by the superposition of differently colored luminescence, 
and a single fluxtube is copied several times in the toroidal direction to form the torus volume. 

Figure 4 shows the distribution of the turbulent particle fluxes along the magnetic field lines, 
where the origin is chosen as the most outboard side of the plasma. 
In order to emphasize the importance of multi-species treatments on the turbulent particle transport, 
we compare the results for the case with multi-ion species (D-T) 
and with the single-ion approximation by $A_{\mr{eff}}=2.5$. 
The ballooning-type distributions appear in all the species of D, T, and $A_{\mr{eff}}$, 
where the outward particle flux has the peak around $z\! =\! 0$. 
It should be noted that the inward particle flux of D, i.e., $\Gamma^{\mr{(local)}}_{\mr{D}}<0$, 
occurs around the inboard side of $z \! \sim \! \pi$, whereas the single-ion case indicating the globally outward flux does not reproduce 
such a flip of the particle-transport direction on the flux surface.        
These results highlight an essential difference between the single-ion approach with the effective ion mass 
and the multi-species simulations, particularly for the particle transport, which will also be seen in Sec. 3.  
%
%
\section{Impacts of D-T ratio and He-ash accumulation on turbulent fluxes}
%
\ \ \ \ In order to further examine the particle-species dependence of turbulent energy and particle fluxes, 
the nonlinear simulation scans with respect to the D-T density ratio and He-ash accumulation have been carried out. 
Here, the accumulations of thermalized He-ash with $n_{\mr{He}}/n_{\mr{e}}$ = \{0\%, 5\%, 10\%\} 
are considered for comparison. 

Figures 5(a) and 5(b) show the dependence of the D-T ratio $n_{\mr{T}}/(n_{\mr{T}}+n_{\mr{D}})$ 
on the mean turbulent energy and particle fluxes for D and T ($Q_{\mr{D,T}}$ and $\Gamma_{\mr{D,T}}$) 
in the cases with and without He-ash, where the ion fluxes are normalized by the electron ones ($Q_{\mr{e}}$ and $\Gamma_{\mr{e}}$). 
Note that both the energy and particle fluxes for electrons are almost constant for all the cases, 
i.e., $Q_{\mr{e}}/Q_{\mr{gB(H)}} \! \simeq \! 110$ and $\Gamma_{\mr{e}}/\Gamma_{\mr{gB(H)}} \! \simeq \! 10$. 
In the figure, the D-T ratio corresponding to the balanced D-T flux, i.e., $Q_{\mr{D}}\! \simeq \! Q_{\mr{T}}$ 
and $\Gamma_{\mr{D}}\! \simeq \! \Gamma_{\mr{T}}$, is shown by the vertical line.   
It is stressed that, for the case with He-ash, the additional transport channel of He-ash 
leads to the reduction in the energy and particle fluxes of D-T ions. 
One also finds that the overall tendency indicating the linear D-T ratio dependence of $Q_{\mr{D}}$ and $Q_{\mr{T}}$ is 
unchanged regardless of He-ash accumulations, 
where the balanced energy flux commonly appears at nearly 50\%-50\% D-T ratio of $n_{\mr{T}}/(n_{\mr{T}}+n_{\mr{D}}) \! = \! 0.5$. 

On the other hand, for the D-T particle fluxes,  
an imbalanced particle transport associated with the nonlinear D-T ratio dependence is revealed, 
where $\Gamma_{\mr{D}}\! \neq \! \Gamma_{\mr{T}}$ even at 50\%-50\% D-T ratio. 
Moreover, the finite He-ash accumulations give rise to not only the reduction in the turbulent flux, 
but also a more significant imbalance of the D-T particle transport. 
%
%
As also shown by the triangle symbols in the figures, one can see that the single-ion approximation with $A_{\mr{eff}}$ 
fails to reproduce the imbalanced transport nature, particularly for the particle transport.     

The difference between the multi-ion species treatments and the single-ion approximation is attributed to the constraint 
by the ambipolar condition for the particle fluxes. 
In the single-ion approximation, $\Gamma_{A_{\mr{eff}}}$  must always be balanced with $\Gamma_{\mr{e}}$, 
so that $(1/2)\Gamma_{A_{\mr{eff}}}/\Gamma_{\mr{e}} \! = \! 1/2$ exactly holds. 
On the other hand, in the multi-ion species case, $\Gamma_{\mr{D}}+\Gamma_{\mr{T}}\! =\! \Gamma_{\mr{e}}$ permits 
the individual values for the ion particle fluxes which can cause the imbalanced particle transport. 
Since there is no such constraint for the thermal flux, the difference between multi-ion and single-ion cases becomes less significant. 

The imbalanced particle transport essentially resulting from multi-ion species effects, which strongly depends on the background D-T ratio 
and the He-ash accumulation, may enable us to develop more efficient D/T fueling and the density-profile controls in burning plasmas.
Then the D-T density ratio providing the balanced flux is regarded as a measure for the strength of imbalance 
in the global D-T ratio dependence, i.e., a larger deviation from $n_{\mr{T}}/(n_{\mr{T}}+n_{\mr{D}})\! =\! 0.5$ 
suggests a more significant imbalanced flux. 
%
\section{Gyrokinetic-simulation-based evaluation of steady burning condition}
%
\ \ \ \ \ The present multi-ion species gyrokinetic turbulence simulations with the thermalized He-ash 
can provide quantitative evaluations of the steady burning condition, 
which can never be addressed by the single-ion-species approaches.  
The profile conditions with optimal balance of turbulent energy and particle transport in the steady burning state 
are discussed in this section.  
%


For sustainable fusion reactions, the He-ash should be exhausted from the core region, 
where the ratio of $\tau_{\mr{He}}/\tau_{\mr{E}}$ is the critical parameter as shown in Fig. 1. 
Then, Reiter's steady burning condition\cite{reiter} is represented by 
$\tau_{\mr{He}} < \alpha^{\ast}\tau_{\mr{E}}$, 
where $\alpha^{\ast} \! = \!$ 7 -- 15 is a constant depending on the wall and divertor conditions. 
For simplicity, we ignore the recycling and exhaust effects.  
This relation describes the permissible particle confinement time of He-ash, compared to the energy confinement time.  

To evaluate Reiter's condition in terms of the gyrokinetic turbulence simulations, the expression with turbulent fluxes is useful. 
Introducing the approximations of $\tau_{\mr{He}} \! \simeq \! a^2/D_{\mr{He}}^{\mr{(eff)}} = a^2 / (\Gamma_{\mr{He}}/n_{\mr{He}}L_{n}^{-1})$ 
and $\tau_{\mr{E}} \! \simeq \! a^2/\chi_{\mr{i}}^{\mr{(eff)}} = a^2 / (Q_{\mr{i}}/n_{\mr{i}}T_{\mr{i}}L_{T_{\mr{i}}}^{-1})$, 
Reiter's steady burning condition with the He-ash exhaust and the fuel inward pinch is rewritten as 
%
\begin{eqnarray}
\eta_{\mr{i}}T_{\mr{i}}(n_{\mr{i}}/n_{\mr{He}})\Gamma_{\mr{He}} > Q_{\mr{i}}/\alpha^{\ast}, \\ 
\Gamma_{\mr{He}} > 0, \\
\Gamma_{\mr{D},\mr{T}} < 0, 
\end{eqnarray}   
%
where $Q_{\mr{i}}$ denotes the summation of the ion energy flux, and $\eta_{\mr{i}} \! = \! L_{n}/L_{T_{\mr{i}}}$. 
In the following, we consider $\alpha^{\ast} \! = \! 7$ as a typical case. 

In order to identify the optimal profile regimes, the density-gradient($R_{\mr{ax}}/L_{n}$) scans of the turbulent energy and particle transport 
are performed, where the case for D-T-He($n_{\mr{He}}/n_{\mr{e}} \! =\! 0.1$) plasma is examined. 
As shown in Fig. 6(a), the density gradient dependence of $T_{\mr{i}}\Gamma_{\mr{D}}$ and $T_{\mr{i}}\Gamma_{\mr{T}}$ 
gives rise to a threshold of the inward D-T fluxes with $\Gamma_{\mr{D,T}} \! \leqslant \! 0$ shown by the vertical dashed line. 
It is expected that the inward pinch related to the off-diagonal contributions from $R_{\mr{ax}}/L_{T_{\mr{i,e}}}$ 
prevails over the outward diffusion from $R_{\mr{ax}}/L_{n}$ in the D-T particle fluxes of $\Gamma_{\mr{D}}$ and $\Gamma_{\mr{T}}$. 
%
Indeed, the wavenumber spectra of $\eta_{\mtr{i}}T_{\mtr{i}}\Gamma_{\mr{He}}$ and $T_{\mtr{i}}\Gamma_{\mr{T}}$ shown in Fig. 6(b) 
indicate the nonuniform changes, where the spectra do not change globally(like a flip of spectrum), 
but the components of the inward tritium flux in the relatively higher wavenumber modes gradually increase 
as $R_{\mr{ax}}/L_{n}$ decreases, 
in contrast to the weak $R_{\mr{ax}}/L_{n}$ dependence of $\Gamma_{\mr{He}}>0$. 
This can also be attributed to the change of the turbulence drive by the ITG and TEM instabilities.

Then, the profile regime of $R_{\mr{ax}}/L_n \leq 1.2$[hatched region in Fig. 6(a)] 
to satisfy the above steady burning condition of Eqs. (11)--(13) with $\alpha^{\ast} \! = \! 7$ is clearly identified. 
Note also that we see $\Gamma_{\mr{e}} \! \simeq \! 0$ around $R_{\mr{ax}}/L_n \! = \! 1$, 
which is necessary for the steady electron density profile. 

To further explore the other possible profile regimes to satisfy the steady burning condition, 
$R_{\mr{ax}}/L_{T_{\mr{e}}}$ and $R_{\mr{ax}}/L_{T_{\mr{i}}}$ scans are performed as shown in Figs. 7(a) and 7(b), respectively, 
where $R_{\mr{ax}}/L_n \! =\! 1$ is fixed, and the encircled symbols in each figure correspond to the previous one shown 
in the hatched region of Fig. 6(a). 
It is found that the larger $R_{\mr{ax}}/L_{T_{\mr{e}}}$ tends to produce the inward particle fluxes for D, T, and electrons, 
while the opposite dependence appears for $R_{\mr{ax}}/L_{T_{\mr{i}}}$. 
Thus, another steady burning profile regime with steeper temperature gradients, but still keeping 
$\Gamma_{\rm{e}} \! \simeq \! 0$, is anticipated for larger $R_{\rm{ax}}/L_{T_{\rm{e}}}$ and $R_{\rm{ax}}/L_{T_{\rm{i}}}$. 
Actually, it has been confirmed that the case with 
\{$R_{\rm{ax}}/L_{n}=1$, $R_{\rm{ax}}/L_{T_{\rm{e}}}=10$, $R_{\rm{ax}}/L_{T_{\rm{i}}}=8$\} 
also satisfies the steady burning condition with 
$\eta_{\mr{i}}T_{\mr{i}}(n_{\mr{i}}/n_{\mr{He}})\Gamma_{\mr{He}}/Q_{\mr{i}} \! = \! 0.496 \! > \! 1/\alpha^{\ast}$, 
$T_{\mr{i}}\Gamma_{\mr{D}} \! = \! -0.291 \! < \! 0$, and  $T_{\mr{i}}\Gamma_{\mr{T}} \! = \! -2.65 \! < \! 0$, 
as well as \{$R_{\rm{ax}}/L_{n}=1$, $R_{\rm{ax}}/L_{T_{\rm{e}}}=8$, $R_{\rm{ax}}/L_{T_{\rm{i}}}=6$\} shown in Fig. 6(a).  

Finally, the impacts of the zonal-flow generation and nonthermal He-ash on the optimal profile regime are discussed. 
As shown in Fig. 8(a), for the case with numerically suppressed zonal flows, the inward D-T particle transport no longer appears 
while the other two conditions of Eqs. (11) and (12) still hold. 
It is thus considered that the zonal flows play an important role in determining the direction of the D-T particle transport.    

Moreover, the results with the nonthermal He-ash are shown in Fig. 8(b), where $T_{\mr{He}}/T_{\mr{i}}=2.5$, 
and $T_{\mr{i}} \! = \! T_{\mr{D}} \! = \! T_{\mr{T}}$. 
Since the He-ash with various temperatures is expected due to the continuous thermalization process for the energetic $\alpha$-particles, 
the impact of He-ash with $T_{\mr{He}}/T_{\mr{i}} \! \neq \! 1$\cite{nakata_pfr2022} on the turbulent fluxes 
is examined here. 
One clearly finds that the significant decrease of the He-ash particle flux leads to the violation of Eq. (11) for $R_{\mr{ax}}/L_n \leqslant 2$. 
Note also that the inward He-ash transport appears for $R_{\mr{ax}}/L_{n} \! \leqslant \! 1.5$. 
This suggests the importance of treating the thermal and nonthermal He-ash simultaneously 
in evaluations of the steady burning condition in realistic situations. 
%
%
\section{Summary}
\ \ \ \ In this study, the ITG-TEM driven turbulent transport in ITER-like D-T-He plasma 
is investigated by means of multi-species gyrokinetic Vlasov simulations. 
It is found that the accumulation of the thermalized He-ash has significant impacts on the turbulent particle fluxes, 
which lead to imbalanced D-T transport. 
Nonlinear scans with the various density and temperature gradient parameters reveal 
the off-diagonal contributions to the deuterium and tritium particle transport, and identify several profile regimes 
that satisfy the steady burning condition, 
where the relatively flat density profile of $R_{\mr{ax}}/L_n \! \sim \! 1$ is expected to be suitable 
for the fuel inward pinch. 
The present numerical studies clearly highlight the importance of the multi-species approach, 
which the effective single-ion calculations cannot replace.

Then the new findings on the D-T ratio impact and on the profile dependence 
for the steady burning condition including the zonal-flow and nonthermal He-ash effects 
contribute to deeper understanding of the multi-species transport processes in burning plasmas. 
This work proposed a novel gyrokinetic-simulation-based analysis of the steady burning condition, 
but more extended investigations including the radial dependence and the high-energy $\alpha$ particles, 
which remain as future work, are useful for the integrated predictive simulations of kinetic profiles in ITER and DEMO. 
%
%
\section*{Acknowledgement}
\ \ \ \ The Authors thank Dr. Masanori Nunami for fruitful discussions. 
This work is supported by the MEXT Japan, Grant Nos. 17K14899, 17K07001, 22K03574, 
in part by the NIFS collaborative Research Programs (NIFS22KIST017, NIFS22KIST018), 
and in part by JST, PRESTO Grant Number JPMJPR21O7, 
and in part by PLADyS, JSPS Core-to-Core Program, 
and in part by Program for Promoting Researches on the Supercomputer Fugaku(Exploration of burning plasma confinement physics, JPMXP1020200103). 

Numerical simulations were performed by Plasma Simulator at NIFS, and JFRS-1 at IFERC-CSC. 
%
%

%
%
\begin{figure}[b]
\begin{center}
 \includegraphics[scale=1.15]{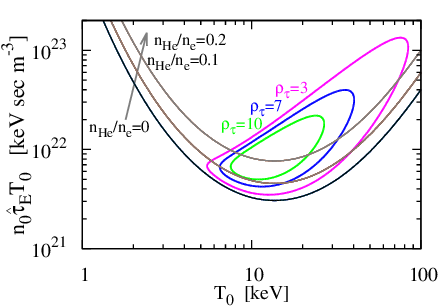}\\
\caption{
Ignition curves influenced by He-ash accumulation($n_{\mr{He}}/n_{\mr{e}}$) 
and the steady burning curves for $\rho_{\tau}\! := \!\tau_{\mr{p}}/\tau_{\mr{E}}$, where $\tau_{\mr{p}} \! \simeq \! \tau_{\mr{He}}$. 
See also Ref. 14 and 15. 
}
\end{center}
\end{figure} 
\begin{figure}[t]
\begin{center}
 \includegraphics[scale=1.0]{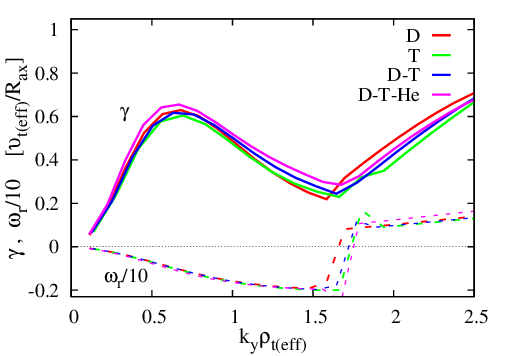}\\
\caption{
Normalized linear growth rate and mode frequency of ITG($\omega_r<0$) and TEM($\omega_r>0$) instabilities 
in pure-D, pure-T, D-T($n_{\mr{D}}\! =\! n_{\mr{T}} \! =\! 0.5n_{\mr{e}}$), 
and D-T-He($n_{\mr{D}}\! =\! n_{\mr{T}} \! =\! 0.4n_{\mr{e}}$, $n_{\mr{He}}\! =\! 0.1n_{\mr{e}}$) plasmas. 
}
\end{center}
\end{figure} 
\begin{figure}[t]
\begin{center}
 \includegraphics[scale=0.46]{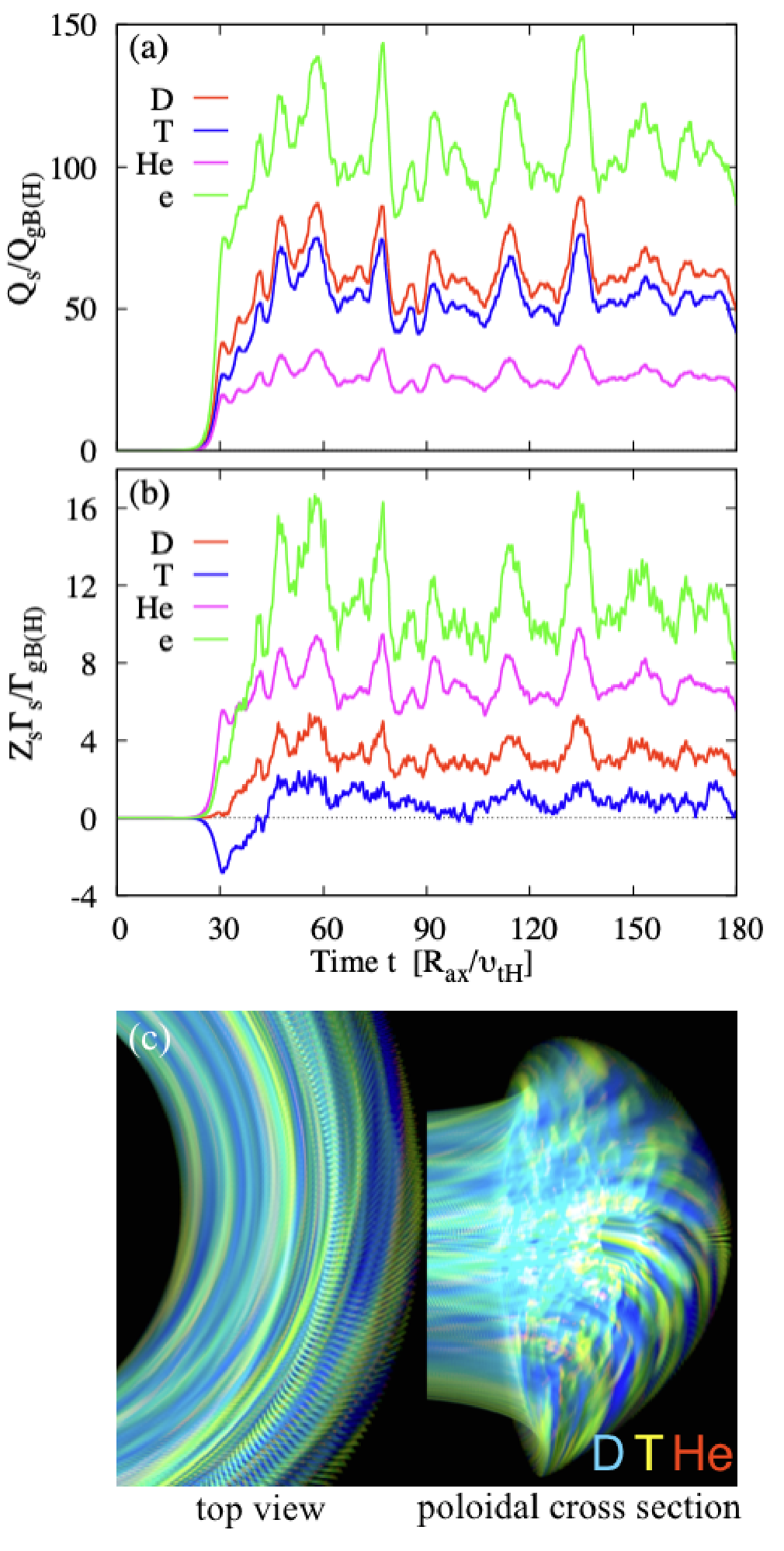}\\
\caption{
Time evolution of (a)turbulent energy flux 
and (b)particle flux in the ITG-TEM turbulence of D-T-He plasma. 
(c)Temperature fluctuations at $t = 65.4R_{\mtr{ax}}/v_{\mtr{tH}}$ of D(blue), T(yellow), and He(red). 
The fluctuations are simultaneously visualized by the superposition of differently colored luminescence. 
}
\end{center}
\end{figure} 
\begin{figure}[t]
\begin{center}
 \includegraphics[scale=1.0]{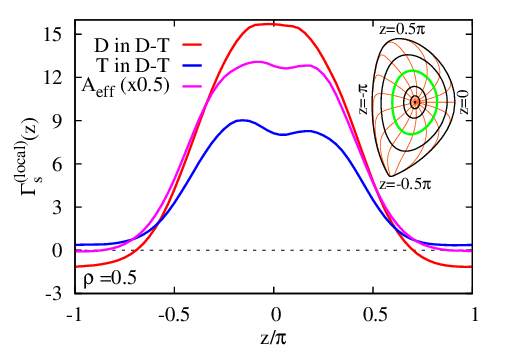}\\
\caption{
Distributions of the turbulent particle fluxes along the field lines in the D-T plasma and the single-ion approximated 
plasma with the effective ion mass $A_{\mr{eff}}$. The flux surface of $\rho=0.5$ is displayed by the curve in green. 
}
\end{center}
\end{figure} 
\begin{figure}[t]
\begin{center}
 \includegraphics[scale=1.0]{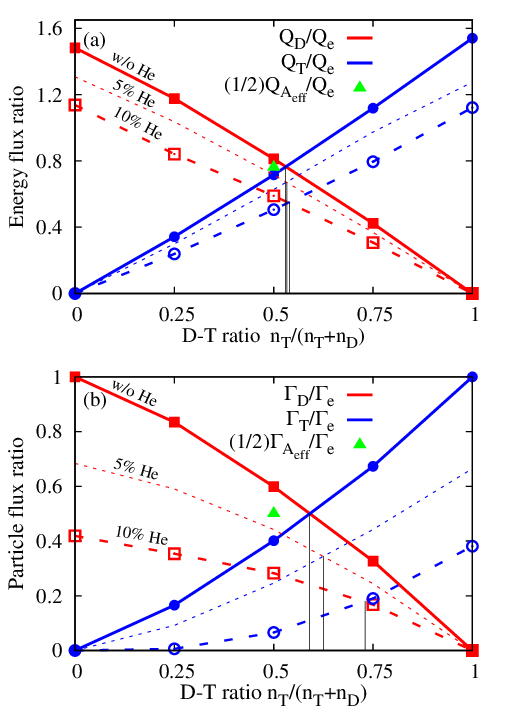}\\
\caption{
D-T ratio dependencies of the mean (a)energy and (b)particle fluxes of D and T 
for He-ash accumulations of $n_{\mr{He}}/n_{\mr{e}}$ = \{0\%(solid), 5\%(dotted), and 10\%(dashed)\}, 
where the ion fluxes are normalized by the electron ones.
The D-T ratio corresponding to the balanced flux is shown by the vertical line.  
}
\end{center}
\end{figure} 
\begin{figure}[t]
\begin{center}
 \includegraphics[scale=0.38]{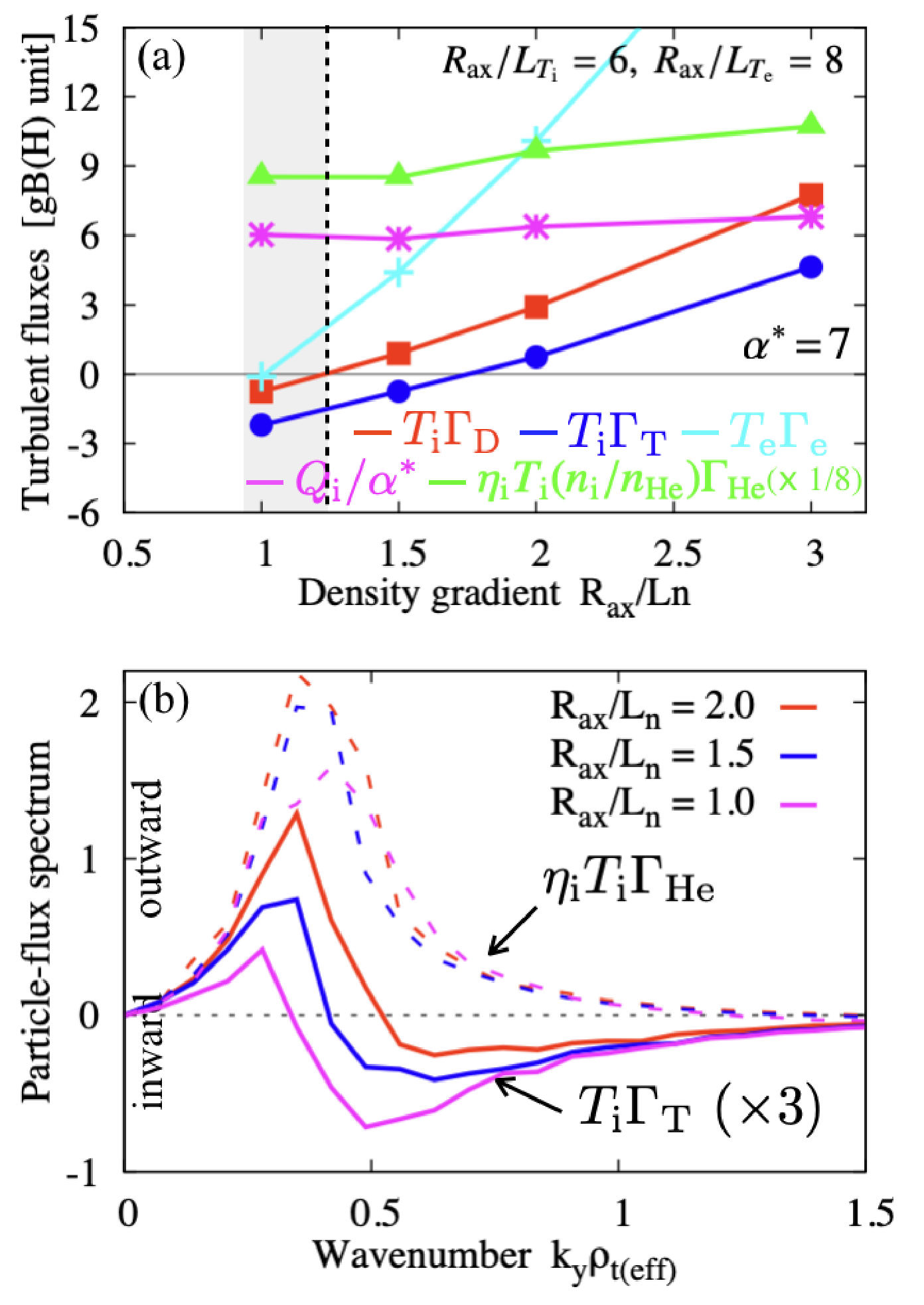}\\
\caption{
(a)Density gradient dependence of turbulent fluxes, 
where $\eta_{\mr{i}}T_{\mr{i}}(n_{\mr{i}}/n_{\mr{He}})\Gamma_{\mr{He}}$ is scaled by the factor of 1/8 for visibility. 
The steady burning condition with $\alpha^{\ast} \! =\! 7$ is satisfied in $R_{\rm{ax}}/L_n \leq 1.27$(grey-hatched region).
(b)Wavenumber spectra of the T(solid) and He(dashed) particle fluxes. 
}
\end{center} 
\end{figure} 
\begin{figure}[t]
\begin{center}
 \includegraphics[scale=0.38]{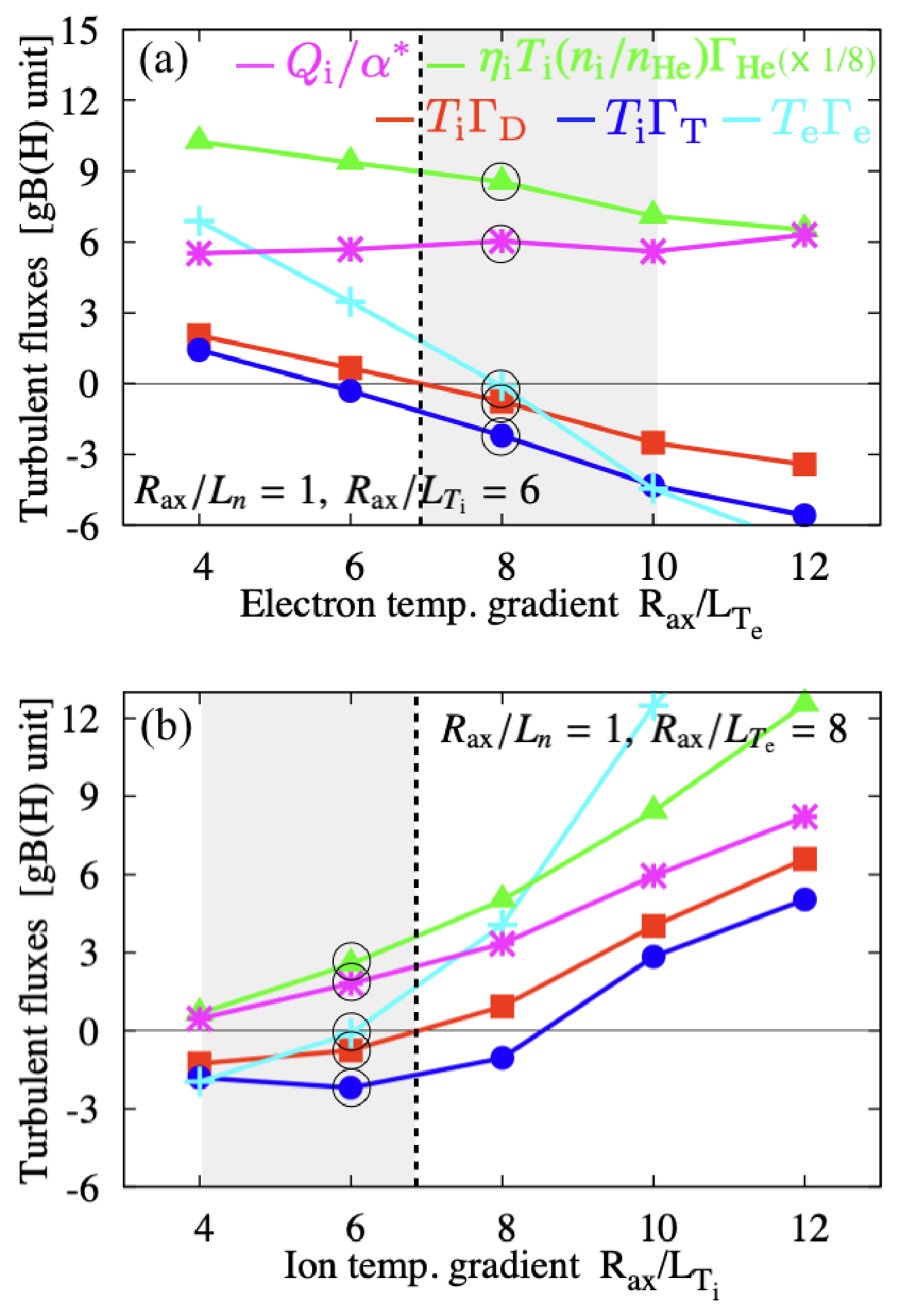}\\
\caption{
Dependence of (a)electron and (b)ion temperature gradient on turbulent particle and ion energy fluxes, 
where $\eta_{\mr{i}}T_{\mr{i}}(n_{\mr{i}}/n_{\mr{He}})\Gamma_{\mr{He}}$ is scaled by the factor of 1/8 for visibility. 
The results encircled in black correspond to the profile regime that satisfies the steady burning condition in Fig. 6(a).  
}
\end{center}
\end{figure} 
\begin{figure}[t]
\begin{center}
 \includegraphics[scale=0.38]{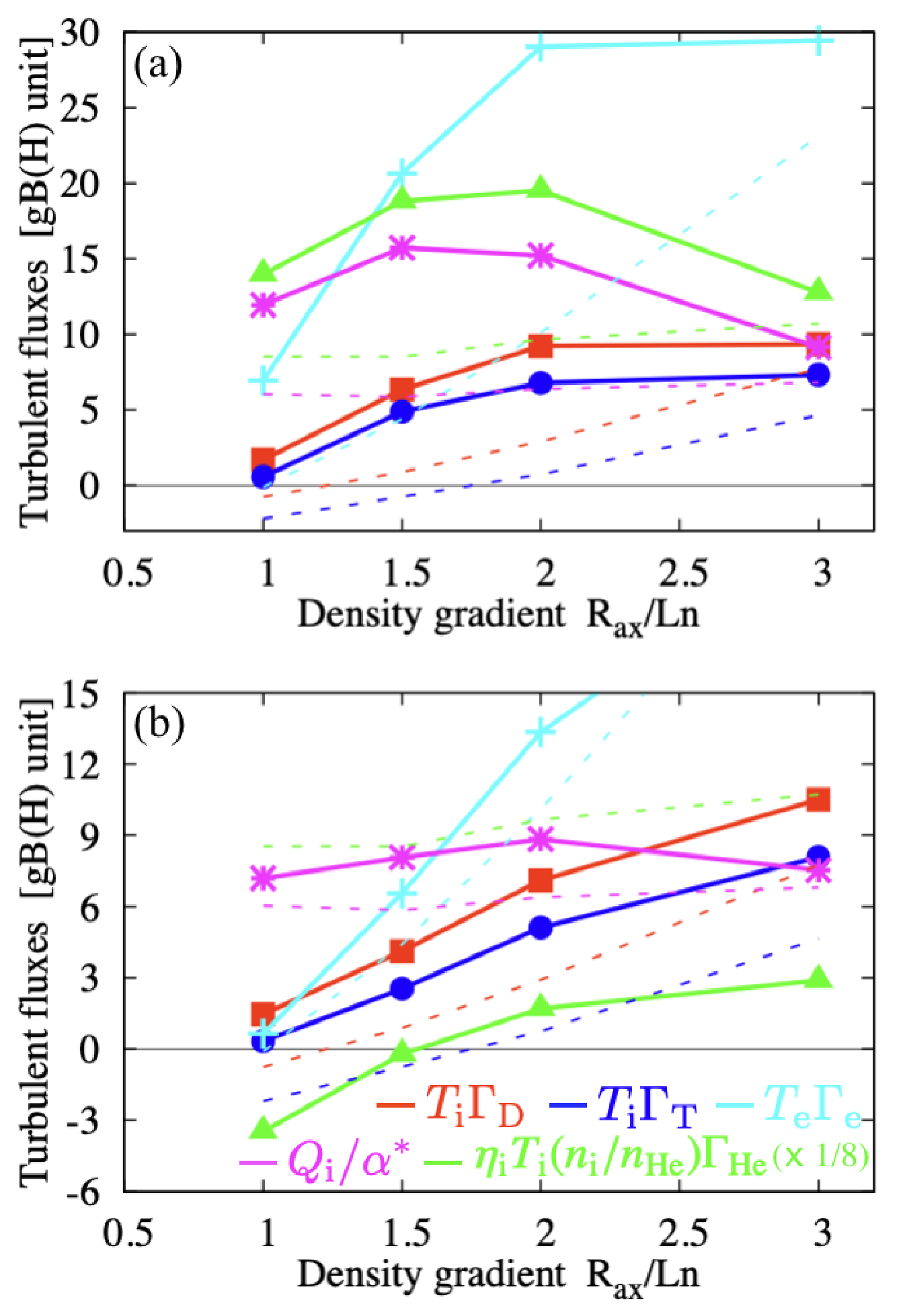}\\
\caption{
Density gradient dependence of turbulent particle and ion energy fluxes in the cases (a)without the zonal flow 
and (b)with the nonthermal He-ash with $T_{\mr{He}}/T_{\mr{i}}=2.5$, 
where $\eta_{\mr{i}}T_{\mr{i}}(n_{\mr{i}}/n_{\mr{He}})\Gamma_{\mr{He}}$ is scaled by the factor of 1/8 for visibility. 
For comparison, the results shown in Fig. 6(a) are displayed by the dashed curves. 
}
\end{center}
\end{figure} 


\begin{thebibliography}{100}
\bibitem{estrada1} C. Estrada-Mila et al., Phys. Plasmas 12, 022305 (2005)
\bibitem{estrada2} C. Estrada-Mila et al., Phys. Plasmas 13, 112303 (2006)
\bibitem{nakata_prl} M. Nakata et al., Phys. Rev. Lett. 118, 165002 (2017)
\bibitem{belli} E. A. Belli et al., Phys. Rev. Lett. 125, 015001 (2020)
\bibitem{Bourdelle} C. Bourdelle et al., Nucl. Fusion 58, 076028 (2018)
\bibitem{angioni1} C. Angioni et al., Nucl. Fusion 49, 055013 (2009)
\bibitem{angioni2} C. Angioni, Phys. Plasmas 22, 102501 (2015)
\bibitem{nakata_pfr2022} M. Nakata et al., Plasma Fusion Res., in press
\bibitem{maeyama} S. Maeayama et al., submitted
\bibitem{ida1} K. Ida et al., Phys. Rev. Lett. 124, 025002 (2020)
\bibitem{ida2} K. Ida et al., Nucl. Fusion 61, 016012 (2021)
\bibitem{urano} H. Urano et al., Nucl. Fusion 55, 033010 (2015)
\bibitem{Mailloux} J. Mailloux et al., Nucl. Fusion, in press
\bibitem{reiter} D. Reiter et al., Nucl. Fusion 30 2141 (1990)
\bibitem{rebhan} E. Rebhan et al., Nucl. Fusion 36 264 (1996)
\bibitem{nakata_pfr2014} M. Nakata et al., Plasma Fusion Res. 9, 1403029 (2014)
\bibitem{nakata_cpc2015} M. Nakata et al., Comput. Phys. Commun. 197, 61 (2015)
\bibitem{nakata_nf2016} M. Nakata et al., Nucl. Fusion 56, 080016 (2016)
\bibitem{gkv} T. -H. Watanabe et al., Nucl. Fusion 46, 24 (2006)
\bibitem{sugama_pop2009} H. Sugama et al., Phys. Plasmas 16, 112502 (2009)
\bibitem{nakata_pop2012} M. Nakata et al., Phys. Plasmas 19, 022303 (2012)

\end{thebibliography}
\end{document}